\documentclass[fleqn,twoside]{article}
\usepackage{espcrc2}

\usepackage{graphicx}
\usepackage[figuresright]{rotating}

\usepackage{cite}

\def\NPPS{Nucl. Phys. B (Proc. Suppl.)}
\def\PL{Phys. Lett.}

\def\PR{Phys. Rev.}

\def\MP{Int. J. Mod. Phys.}

\newcommand{\be}{\begin{equation}}
\newcommand{\ee}{\end{equation}}
\newcommand{\no}{\nonumber}

\newcommand{\ba}{\begin{array}{c}}
\newcommand{\bat}{\begin{array}{cc}}
\newcommand{\ea}{\end{array}}
\newcommand{\beqn}{\begin{eqnarray}}
\newcommand{\eeqn}{\end{eqnarray}}

\newcommand{\bi}{\begin{itemize}}
\newcommand{\ei}{\end{itemize}}

\newcommand{\Br}{\mathrm{Br}}

\newcommand{\AmS}{{\protect\the\textfont2
  A\kern-.1667em\lower.5ex\hbox{M}\kern-.125emS}}

\title{Tau Physics: Theory Overview}

\author{A. Pich\address{Departament de F\'{\i}sica Te\`orica, IFIC,
Univ. Val\`encia--CSIC, Apt. 22085, E-46071 Val\`encia, Spain}
}

\begin{document}

\begin{abstract}
The present status of some selected topics on $\tau$ physics is presented:
charged-current universality tests, bounds on lepton-flavour violation,
the determination of $\alpha_s$ from the inclusive $\tau$ hadronic width,
the measurement of $|V_{us}|$ through the Cabibbo-suppressed decays
of the $\tau$, and the theoretical description of the $\tau\to\nu_\tau K \pi$ spectrum.
\vspace{1pc}
\end{abstract}

\maketitle

\section{Introduction}

The known leptons provide clean probes to precisely test
the Standard Model and search for signals of new dynamics. The
electroweak gauge structure has been successfully tested at the
0.1\% to 1\% level, confirming the Standard Model framework
\cite{CERN07}.
Moreover, the hadronic $\tau$ decays turn out to be a beautiful
laboratory for studying strong interaction effects at low energies \cite{taurev06}.
The $\tau$ is the only known lepton massive enough to decay into hadrons.
Its semileptonic decays are then ideally suited for studying the
hadronic weak currents. 
Accurate determinations of the QCD coupling, $|V_{us}|$ and the
strange quark mass have been obtained with $\tau$ decay data.

The huge statistics accumulated at the B Factories allow to explore
lepton-flavour-violating $\tau$ decay modes with increased
sensitivities beyond $10^{-7}$, which could be further pushed down
to few $10^{-9}$ at future facilities. Moreover, BESIII will soon
start taking data at threshold, providing
complementary information on the $\tau$,
such as an improved mass measurement.
Thus, $\tau$ physics is entering a new era, full of interesting
possibilities and with a high potential for new discoveries.

\section{Tests on charged-current universality}

\begin{table}[t] 
\centering \caption{Present constraints on $|g_l/g_{l'}|$.}
\label{tab:ccuniv}
\vspace{0.2cm}
\renewcommand{\tabcolsep}{1.1pc} 
\renewcommand{\arraystretch}{1.15} 
\begin{tabular}{lc}
\hline & $|g_\mu/g_e|$ \\ \hline
$B_{\tau\to\mu}/B_{\tau\to e}$ & $1.0000\pm 0.0020$ \\
$B_{\pi\to \mu}/B_{\pi\to e}$ &  $1.0021\pm 0.0016$ \\ 
$B_{K\to \mu}/B_{K\to e}$ & $1.004\pm 0.007$ \\        
$B_{K\to \pi\mu}/B_{K\to\pi e}$ & $1.002\pm 0.002$ \\
$B_{W\to\mu}/B_{W\to e}$  & $0.997\pm 0.010$ \\
\hline\hline & $|g_\tau/g_\mu|$  \\ \hline
$B_{\tau\to e}\,\tau_\mu/\tau_\tau$ & $1.0006\pm 0.0022$ \\
$\Gamma_{\tau\to\pi}/\Gamma_{\pi\to\mu}$ &  $0.996\pm 0.005$ \\
$\Gamma_{\tau\to K}/\Gamma_{K\to\mu}$ & $0.979\pm 0.017$ \\
$B_{W\to\tau}/B_{W\to\mu}$  & $1.039\pm 0.013$
\\ \hline\hline
& $|g_\tau/g_e|$  \\ \hline
$B_{\tau\to\mu}\,\tau_\mu/\tau_\tau$ & $1.0005\pm 0.0023$ \\
$B_{W\to\tau}/B_{W\to e}$  & $1.036\pm 0.014$
\\ \hline
\end{tabular}
\end{table}
%

In the Standard Model all lepton doublets have identical couplings
to the $W$ boson. Comparing the measured decay widths of leptonic or
semileptonic decays which only differ in the lepton flavour, one can
test experimentally that the $W$ interaction is indeed the same,
i.e. that \ $g_e = g_\mu = g_\tau \equiv g\, $. As shown in
Table~\ref{tab:ccuniv}, the present data verify the universality of
the leptonic charged-current couplings to the 0.2\% level.

\begin{table*}[tbh]
\caption{Best published limits (90\% CL) on lepton-flavour-violating
decays \cite{PDG,LFVbabar,LFVbelle}.}
\label{table:LFV}\vspace{0.2cm}
\renewcommand{\tabcolsep}{0.5pc} 
\renewcommand{\arraystretch}{1.2} 
\begin{tabular}{@{}llllllllll@{}}
\hline
\multicolumn{6}{@{}l}{$\rm{Br}(\mu^-\to X^-)\cdot 10^{12}$}\\
 $e^-\gamma$ & $12$ &
 $e^-2\gamma$ & $72$ &
 $e^-e^-e^+$ & $\phantom{1}1.0$ &&&&
 \\
 \hline\hline
 \multicolumn{6}{@{}l}{$\rm{Br}(\tau^-\to X^-)\cdot 10^{8}$}\\
 $e^-\gamma$ & $11$ &
 $e^-e^+e^-$ & $\phantom{1}3.6$ &
 $e^-\mu^+\mu^-$ & $\phantom{1}3.7$ &
 $e^-e^-\mu^+$ & $\phantom{1}2.0$ &
 $e^-\pi^0$ & $\phantom{1}8.0$
 \\
 $\mu^-\gamma$ & $\phantom{1}4.5\quad\;$ &
 $\mu^-e^+e^-$ & $\phantom{1}2.7\quad\;$ &
 $\mu^-\mu^+\mu^-$ & $\phantom{1}3.2\quad\;$ &
 $\mu^-\mu^-e^+$ & $\phantom{1}2.3\quad\;$ &
 $\mu^-\pi^0$ & $11$
 \\
 $e^-\eta$ & $\phantom{1}9.2$ &
 $e^-\eta'$ & $16$ &
 $e^-\rho^0$ & $\phantom{1}6.3$ &
 $e^-\omega$ & $11$ &
 $e^-\phi$ & $\phantom{1}7.3$
 \\
 $\mu^-\eta$ & $\phantom{1}6.5$ &
 $\mu^-\eta'$ & $13$ &
 $\mu^-\rho^0$ & $\phantom{1}6.8$ &
 $\mu^-\omega$ & $\phantom{1}8.9$ &
 $\mu^-\phi$ & $13$
 \\
 $e^-K_S$ & $\phantom{1}5.6$ &
 $e^-K^{* 0}$ & $\phantom{1}7.8$ &
 $e^-\bar K^{* 0}$ & $\phantom{1}7.7$ &
 $e^-K^+\pi^-$ & $16$ &
 $e^-\pi^+K^-$ & $32$
 \\
 $\mu^-K_S$ & $\phantom{1}4.9$ &
 $\mu^-K^{*0}$ & $\phantom{1}5.9$ &
 $\mu^-\bar K^{*0}$ & $10$ &
 $\mu^-K^+\pi^-$ & $32$ &
 $\mu^-\pi^+K^-$ & $26$
 \\
 $e^-K^+K^-$ & $14$ &
 $e^-\pi^+\pi^-$ & $12$ &
 $e^+\pi^-\pi^-$ & $20$ &
 $e^+K^-K^-$ & $15$ &
 $e^+\pi^-K^-$ & $18$
 \\
 $\mu^-K^+K^-$ & $25$ &
 $\mu^-\pi^+\pi^-$ & $29$ &
 $\mu^+\pi^-\pi^-$ & $7$ &
 $\mu^+K^-K^-$ & $44$ &
 $\mu^+\pi^-K^-$ & $22$
 \\
 $\bar\Lambda\pi^-$ & $14$ &
 $\Lambda\pi^-$ & $\phantom{1}7.2$ &
 \\ \hline 
\end{tabular}\end{table*}

The $\tau$ leptonic branching fractions and the $\tau$ lifetime are
known with a precision of $0.3\%$ \cite{taurev06}. A slightly improved lifetime
measurement could be expected from BaBar and Belle \cite{Lusiani}.
For comparison, the $\mu$ lifetime is already known with an accuracy of
$10^{-5}$, which should be further improved to $10^{-6}$ by the
MuLan \cite{MuLan:07} and FAST \cite{FAST:08} experiments at PSI.
The universality tests require also a good determination of
$m_\tau^5$, which is only known to the $0.06\%$ level \cite{PDG}. Two new
measurements of the $\tau$ mass have been published recently:
$$
m_\tau =\left\{ \begin{array}{lr} 1776.61\pm 0.13\pm
0.35~\mathrm{MeV}
&\; [\mathrm{Belle}],\\[10pt]
1776.81\, {}^{+\, 0.25}_{-\, 0.23} \pm 0.15~\mathrm{MeV} &\;
[\mathrm{KEDR}]. \ea\right.
$$
Belle \cite{BelleTauMass} has made a pseudomass analysis of
$\tau\to\nu_\tau 3\pi$ decays, while KEDR \cite{KEDR} measures the
$\tau^+\tau^-$ threshold production, taking advantage of a precise
energy calibration through the resonance depolarization method. In
both cases the achieved precision is getting close to the previous
BES-dominated value, $m_\tau = 1776.99\, {}^{+\, 0.29}_{-\, 0.26}$
\cite{PDG}. KEDR aims to obtain a final accuracy of 0.15 MeV. A
precision better than 0.05 MeV should be easily achieved at BESIII
\cite{MO}, through a detailed analysis of
$\sigma(e^+e^-\to\tau^+\tau^-)$ at threshold \cite{Pedro,Voloshin,SV:94}.

Table~\ref{tab:ccuniv} shows also the contraints obtained from
pion \cite{PDG} and kaon decays \cite{Kl3}, applying the
recently calculated radiative corrections at NLO in chiral perturbation
theory \cite{CR:07,CKNRT:02}. The accuracy achieved with $K_{l3}$ data is
already comparable to the one obtained from $\tau$ or $\pi_{l2}$ decays.

Owing to the limited statistics available, the decays $W^-\to l^-\nu_l$
only test universality at the 1\% level. At present,
$\Br(W\to\nu_\tau\tau)$ is $2.1\,\sigma/2.7\,\sigma$ larger than
$\Br(W\to \nu_e e / \nu_\mu\mu)$ \cite{LEPEWWG}. The stringent limits on
$|g_\tau/g_{e,\mu}|$ from $W$-mediated decays make unlikely that
this is a real effect.

\section{Lepton-flavour violating decays}

We have now clear experimental evidence that neutrinos are
massive particles and there is mixing in the lepton sector.
The smallness of neutrino masses implies a strong suppression of
neutrinoless lepton-flavour-violating processes, which can be
avoided in models with sources of lepton flavour violation
not related to $m_{\nu_i}$.
The scale of the flavour-violating new-physics interactions can be
constrained imposing the requirement of a viable leptogenesis.
Recent studies within different new-physics scenarios find interesting
correlations between $\mu$ and $\tau$ lepton-flavour-violating decays,
with $\mu\to e\gamma$ often expected to be close to the present exclusion limit
\cite{CIP:07,GCIW:07,AAHT:06,AH:06,BBDPT:07,CMPSVV:07,PA:06}.

The B Factories are pushing the experimental
limits on neutrinoless lepton-flavour-violating
$\tau$ decays beyond the $10^{-7}$ level \cite{LFVbabar,LFVbelle},
increasing in a drastic way the sensitivity to new physics scales.
Future experiments could push further some limits to the $10^{-9}$
level \cite{SuperB}, allowing to explore interesting and
totally unknown phenomena. Complementary information will be
provided by the MEG experiment, which will search for $\mu^+\to
e^+\gamma$ events with a sensitivity of $10^{-13}$ \cite{MEG}.
There are also ongoing projects at J-PARC aiming to study $\mu\to e$
conversions in muonic atoms, at the $10^{-16}$ \cite{PRISM-I}
or even $10^{-18}$ \cite{PRISM-II} level.

\section{The inclusive hadronic width of the tau}

The inclusive character of the total $\tau$ hadronic width renders
possible an accurate calculation of the ratio
\cite{BR:88,NP:88,BNP:92,LDP:92a,QCD:94}
%
%
%
$$
 R_\tau \equiv { \Gamma [\tau^- \to \nu_\tau
 \,\mathrm{hadrons}] \over \Gamma [\tau^- \to \nu_\tau e^-
 {\bar \nu}_e] } \, =\,
 R_{\tau,V} + R_{\tau,A} + R_{\tau,S}\, .
$$ 
Using analyticity constraints and the Operator Product Expansion,
one can separately compute the contributions associated with
specific quark currents:
$R_{\tau,V}$ and $R_{\tau,A}$ correspond to the Cabibbo-allowed
decays through the vector and axial-vector currents, while
$R_{\tau,S}$ contains the remaining Cabibbo-suppressed
contributions.

The combination $R_{\tau,V+A}$
can be written as \cite{BNP:92}
\begin{equation}\label{eq:Rv+a}
 R_{\tau,V+A} \, =\, N_C\, |V_{ud}|^2\, S_{\mathrm{EW}} \left\{ 1 +
 \delta_{\mathrm{P}} + \delta_{\mathrm{NP}} \right\} ,
\end{equation}
where $N_C=3$ is the number of quark colours
and $S_{\mathrm{EW}}=1.0201\pm 0.0003$ contains the
electroweak radiative corrections \cite{MS:88,BL:90,ER:02}.
The dominant correction ($\sim 20\%$) is the perturbative QCD
contribution $\delta_{\mathrm{P}}$, which is already known to
$O(\alpha_s^4)$ \cite{BNP:92,BChK:08} and includes a resummation of the most
important higher-order effects \cite{LDP:92a,PI:92}.

%
\begin{figure}[tbh]
\label{fig:alpha_s} \centering
\includegraphics[width=7.6cm]{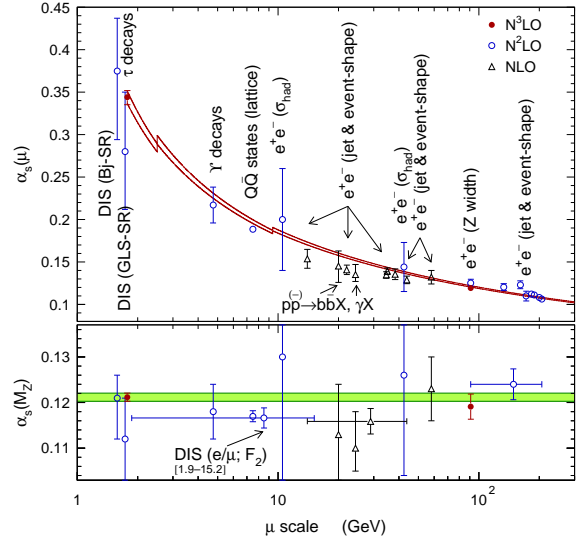}
\vspace{-0.9cm}
\caption{Measured values of $\alpha_s$ at different
scales. The curves show the energy dependence predicted by QCD,
using $\alpha_s(m_\tau)$ as input. The corresponding extrapolated
$\alpha_s(M_Z)$ values are shown at the bottom, where the shaded
band displays the $\tau$ decay result within errors \cite{DHZ:05}.}
\end{figure}

Non-perturbative contributions are suppressed by six powers of the
$\tau$ mass \cite{BNP:92} and, therefore, are very small. Their
numerical size has been determined from the invariant-mass
distribution of the final hadrons in $\tau$ decay, through the study
of weighted integrals \cite{LDP:92b},
\begin{equation}
 R_{\tau}^{kl} \,\equiv\, \int_0^{m_\tau^2} ds\, \left(1 - {s\over
 m_\tau^2}\right)^k\, \left({s\over m_\tau^2}\right)^l\, {d
 R_{\tau}\over ds} \, ,
\end{equation}
which can be calculated theoretically in the same way as $R_{\tau}$.
The predicted suppression \cite{BNP:92} of the non-perturbative
corrections has been confirmed by ALEPH \cite{ALEPH:05}, CLEO
\cite{CLEO:95} and OPAL \cite{OPAL:98}. The most recent analysis
\cite{DHZ:05} gives
\begin{equation}\label{eq:del_np}
 \delta_{\mathrm{NP}} \, =\, 0.0001\pm 0.0017 \, .
\end{equation}

The QCD prediction for $R_{\tau,V+A}$ is then completely dominated
by $\delta_P$; 
non-perturbative effects being
smaller than the perturbative uncertainties from uncalculated
higher-order corrections. The result turns out to be very sensitive
to the value of $\alpha_s(m_\tau)$, allowing for an accurate
determination of the fundamental QCD coupling \cite{NP:88,BNP:92}.
The experimental measurement $R_{\tau,V+A}= 3.478\pm0.011$ implies
\cite{DHZ:05}
\begin{equation}\label{eq:alpha}
 \alpha_s(m_\tau)  \,=\,  0.344\pm
 0.005_{\mathrm{exp}}\pm 0.007_{\mathrm{th}} \, .
\end{equation}

The strong coupling measured at the $\tau$ mass scale is
significantly larger than the values obtained at higher energies.
From the hadronic decays of the $Z$, one gets $\alpha_s(M_Z) =
0.1191\pm 0.0027$ \cite{LEPEWWG}, which differs from $\alpha_s(m_\tau)$
by more than $20\,\sigma$. 
After evolution up to the scale $M_Z$ \cite{Rodrigo:1998zd}, the strong
coupling constant in (\ref{eq:alpha}) decreases to \cite{DHZ:05}
\begin{equation}\label{eq:alpha_z}
 \alpha_s(M_Z)  \, =\,  0.1212\pm 0.0011 \, ,
\end{equation}
in excellent agreement with the direct measurements at the $Z$ peak
and with a better accuracy. The comparison of these two
determinations of $\alpha_s$ in two very different energy regimes, $m_\tau$
and $M_Z$, provides a beautiful test of the predicted running of the
QCD coupling; i.e., a very significant experimental verification of
{\it asymptotic freedom}.

\section{$|V_{us}|$ determination from 
tau decays}

The separate measurement of the $|\Delta S|=0$ and $|\Delta S|=1$
tau decay widths provides a very clean determination of $V_{us}\,$
\cite{GJPPS:05,PI:07b}.
To a first approximation the Cabibbo mixing can be directly obtained
from experimental measurements, without any theoretical input.
Neglecting the small SU(3)-breaking corrections from the $m_s-m_d$
quark-mass difference, one gets:
$$ 
 |V_{us}|^{\mathrm{SU(3)}} =\: |V_{ud}| \:\left(\frac{R_{\tau,S}}{R_{\tau,V+A}}\right)^{1/2}
 =\: 0.210\pm 0.003\, .
$$ 
We have used $|V_{ud}| = 0.97377\pm 0.00027$ \cite{PDG},
$R_\tau = 3.640\pm 0.010$       
and the value $R_{\tau,S}=0.1617\pm 0.0040$ \cite{PI:07b}, which results from the
most recent BaBar \cite{BA:07} and Belle \cite{BE:07} measurements of Cabibbo-suppressed
tau decays \cite{Banerjee}.
The new branching ratios measured by BaBar and Belle are all smaller than the previous
world averages, which translates into a smaller value of $R_S$ and $|V_{us}|$.
For comparison, the previous value $R_{\tau,S}=0.1686\pm 0.0047$ \cite{DHZ:05} resulted in $|V_{us}|^{\mathrm{SU(3)}}=0.215\pm 0.003$.

This rather remarkable determination is only slightly shifted by
the small SU(3)-breaking contributions induced by the strange quark mass.
These corrections can be 
estimated through a QCD analysis of the differences
\cite{GJPPS:05,PI:07b,PP:99,ChDGHPP:01,ChKP:98,KKP:01,MW:06,KM:00,MA:98,BChK:05}
\begin{equation}
 \delta R_\tau^{kl}  \,\equiv\,
 {R_{\tau,V+A}^{kl}\over |V_{ud}|^2} - {R_{\tau,S}^{kl}\over |V_{us}|^2}\, .
\end{equation}
%
The only non-zero contributions are proportional 
to the mass-squared difference $m_s^2-m_d^2$ or to vacuum expectation
values of SU(3)-breaking operators such as $\delta O_4
\equiv \langle 0|m_s\bar s s - m_d\bar d d|0\rangle \approx (-1.4\pm 0.4)
\cdot 10^{-3}\; \mathrm{GeV}^4$ \cite{PP:99,GJPPS:05}. The dimensions of these operators
are compensated by corresponding powers of $m_\tau^2$, which implies a strong
suppression of $\delta R_\tau^{kl}$ \cite{PP:99}:
\beqn\label{eq:dRtau}
 \delta R_\tau^{kl} &\!\!\approx &\!\!  24\, S_{\mathrm{EW}}\; \left\{ {m_s^2(m_\tau)\over m_\tau^2} \,
 \left( 1-\epsilon_d^2\right)\,\Delta_{kl}(\alpha_s)
 \right.\no\\ &&\hskip 1.3cm\left.
 - 2\pi^2\, {\delta O_4\over m_\tau^4} \, Q_{kl}(\alpha_s)\right\}\, ,
\eeqn
where $\epsilon_d\equiv m_d/m_s = 0.053\pm 0.002$ \cite{LE:96}.
The perturbative 
corrections $\Delta_{kl}(\alpha_s)$ and
$Q_{kl}(\alpha_s)$ are known to $O(\alpha_s^3)$ and $O(\alpha_s^2)$,
respectively \cite{PP:99,BChK:05}.

The theoretical analysis of $\delta R_\tau\equiv\delta R_\tau^{00}$
 involves the two-point vector and axial-vector correlators,
which have
transverse ($J=1$) and longitudinal ($J=0$) components.
The longitudinal contribution to $\Delta_{00}(\alpha_s)$ shows a rather
pathological behaviour, with clear signs of being a non-convergent perturbative
series. Fortunately, the corresponding longitudinal contribution to
$\delta R_\tau$ can be estimated phenomenologically with a much better
accuracy, $\delta R_\tau|^{L}\, =\, 0.1544\pm 0.0037$ \cite{GJPPS:05,JOP:06},
because it is dominated by far by the well-known $\tau\to\nu_\tau\pi$
and $\tau\to\nu_\tau K$ contributions. To estimate the remaining transverse
component, one needs an input value for the strange quark mass. Taking the
range
$m_s(m_\tau) = (100\pm 10)\:\mathrm{MeV}$ \
[$m_s(2\:\mathrm{GeV}) = (96\pm 10)\:\mathrm{MeV}$],
which includes the most recent determinations of $m_s$ from QCD sum rules
and lattice QCD \cite{JOP:06},
one gets finally $\delta R_{\tau,th} = 0.216\pm 0.016$, which implies \cite{PI:07b}
\beqn\label{eq:Vus_det}
 |V_{us}| &=& \left(\frac{R_{\tau,S}}{\frac{R_{\tau,V+A}}{|V_{ud}|^2}-\delta
 R_{\tau,\mathrm{th}}}\right)^{1/2}
 \no\\ &=&
 0.2165\pm 0.0026_{\mathrm{\, exp}}\pm 0.0005_{\mathrm{\, th}}\, .
\eeqn
%
A larger central value
($|V_{us}| = 0.2212\pm 0.0031$) 
is obtained with the old world average for $R_{\tau,S}$.

Sizeable changes on the experimental determination of $R_{\tau,S}$ are to be expected from
the full analysis of  the huge BaBar and Belle data samples. In particular, the high-multiplicity
decay modes are not well known at present.
Thus, the result (\ref{eq:Vus_det}) could easily fluctuate in the near future.
However, it is important to realize that the final error of the $V_{us}$ determination from
$\tau$ decay is completely dominated by the experimental uncertainties. If $R_{\tau,S}$
is measured with a 1\% precision, the resulting $V_{us}$ uncertainty will
get reduced to around 0.6\%, i.e. $\pm 0.0013$, making $\tau$ decay the best source of
information about $V_{us}$.

An accurate measurement of the invariant-mass distribution of the final hadrons
could make possible a simultaneous determination
of $V_{us}$ and the strange quark mass, through a correlated analysis of
several weighted differences $\delta R_\tau^{kl}$. The extraction of $m_s$ suffers from
theoretical uncertainties related to the convergence of the perturbative series
$\Delta_{kl}^{L+T}(\alpha_s)$, which makes necessary a better
understanding of these corrections.

%
\begin{figure}[tbh]
\centering
\includegraphics[width=7.6cm]{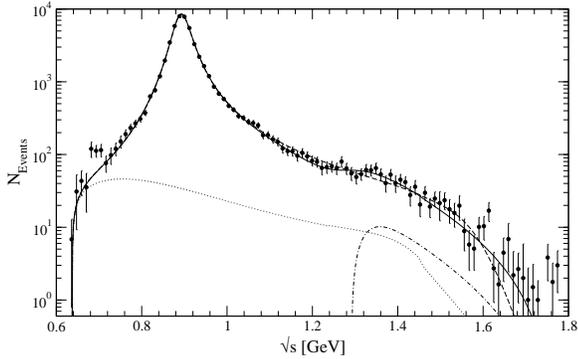}
\vspace{-0.9cm}
\caption{Theoretical description \cite{JPP:08} (solid line) of the Belle
$\tau^-\to\nu_\tau K_S\pi^-$ data \cite{Belle:08}. The $K^{*'}$ (dashed-dotted) and scalar (dotted) contributions are also shown. \label{fig:KpSpectrum}}
\end{figure}

\section{$\tau\to\nu_\tau K\pi$ and $K\to\pi l\bar\nu_l$}

The decays $\tau\to\nu_\tau K\pi$ probe the same hadronic form factors investigated in
$K_{l3}$ processes, but they are sensitive to a much broader range of invariant masses.
A theoretical understanding of the form factors can be achieved, using analyticity, unitarity and some general properties of QCD, such as chiral symmetry and the short-distance asymptotic behaviour. Figure~\ref{fig:KpSpectrum} compares the resulting theoretical description of the $\tau$ decay spectrum \cite{JPP:08}  with the recent Belle measurement \cite{Belle:08}. At low values of $s$ there is clear evidence of the scalar contribution, which was predicted using a careful analysis of $K\pi$ scattering data \cite{JOP:06,JOP:00}. From the measured $\tau$ spectrum one obtains $M_{K^*} = 895.3\pm 0.2$ MeV and $\Gamma_{K^*}=47.5\pm 0.4$ MeV  \cite{JPP:08}. Since the absolute normalization is fixed by
$K_{l3}$ data to be $|V_{us}|\, f_+^{K^0\pi^-}(0)= 0.21664 \pm 0.00048$ \cite{Kl3}, one gets then a theoretical prediction for the branching fraction,
Br$(\tau^-\to\nu_\tau K_S\pi^-) = 0.427 \pm 0.024\%$, in good agreement with the Belle measurement $0.404 \pm 0.013\%$, although slightly larger.

The $\tau$ determination of the vector form factor $f_+^{K\pi}(s)$ provides precise values for its slope and curvature \cite{JPP:08},  $\lambda_+'= (25.2 \pm 0.3)\cdot 10^{-3}$ and $\lambda_+''=
 (12.9 \pm 0.3)\cdot 10^{-4}$, in agreement but more precise than the corresponding $K_{l3}$ measurements~\cite{Kl3}.


\section*{Acknowledgements}
I want to thank the hospitality of the CERN Theory Division.
This work has been supported
by MEC, Spain (grants FPA2007-60323 and
Consolider-Ingenio 2010 CSD2007-00042, CPAN) and by the
EU Contract MRTN-CT-2006-035482 (FLAVIAnet).


\end{document}